\documentclass[lettersize,journal]{IEEEtran}
\usepackage{amsmath,amsfonts,amssymb}
\usepackage{algorithmic}
\usepackage{algorithm}
\usepackage{array}
\usepackage[caption=false,font=normalsize,labelfont=sf,textfont=sf]{subfig}
\usepackage{textcomp}
\usepackage{stfloats}
\usepackage{url}
\usepackage{verbatim}
\usepackage{graphicx}
\usepackage{cite}
\usepackage{tikz}
\usetikzlibrary{matrix,positioning,decorations.pathreplacing}
\usepackage{forest}
\usepackage{booktabs}
\usepackage{orcidlink}

\newcommand{\ex}{\mathrm{e}}
\newcommand{\sign}{\mathrm{s}}
\newcommand{\mant}{\mathrm{m}}

\newcommand{\arr}[1]{\mathbf{#1}}

\begin{document}

\title{\textsc{jz-tree}: GPU friendly neighbour search and friends-of-friends with dual tree walks in JAX plus CUDA}

\author{Jens Stücker \orcidlink{0000-0003-1258-1466}, Oliver Hahn \orcidlink{0000-0001-9440-1152}, Lukas Winkler \orcidlink{0000-0002-6792-6743}, Adrian Gutierrez Adame \orcidlink{0009-0005-0594-9391}, and Thomas Fl\"oss \orcidlink{0000-0002-8245-780X}
\thanks{All authors are at the University of Vienna, Department of Astrophysics, T\"urkenschanzstraße 18, 1180 Vienna, Austria. Oliver Hahn and Thomas Fl\"oss are additionally at University of Vienna, Department of Mathematics, Oskar-Morgenstern-Platz 1, 1090 Vienna, Austria.}
\thanks{E-mail: jens.stuecker@univie.ac.at}
}

\markboth{GPU-Friendly kNN and FoF with Dual Tree Walks in JAX/CUDA}%
{Stücker \MakeLowercase{\textit{et al.}}: GPU-Friendly kNN and FoF with Dual Tree Walks in JAX/CUDA}%


\maketitle

\begin{abstract}
Algorithms based on spatial tree traversal are widely regarded as among the most efficient and flexible approaches for many problems in CPU-based high-performance computing (HPC). However, directly transferring these algorithms to GPU architectures often yields substantially smaller performance gains than expected in light of the high computational throughput of modern GPUs. The branching nature of tree algorithms leads to thread divergence and irregular memory access patterns -- both of which may severely limit GPU performance.

To address these challenges, we propose a Morton (z-order) \emph{plane-based tree hierarchy} that is specifically designed for GPU architectures. The resulting flattened data layout enables efficient dual-tree traversal with collaborative execution across thread groups, leading to highly coalesced memory access patterns. 

Based on this framework we present implementations of two important spatial algorithms -- exact $k$-nearest neighbour search and friends-of-friends (FoF) clustering. For both cases, we observe more than an order-of-magnitude performance improvement over the closest competing GPU libraries for large problem sizes ($N \gtrsim 10^7$), together with strong scaling to distributed multi-GPU systems.

We provide an open-source implementation, \textsc{jz-tree} (\textsc{jax} z-order tree), which serves as a foundation for efficient GPU implementations of a broad class of tree-based algorithms.
\end{abstract}


\section{Introduction}

\IEEEPARstart{H}{igh}-performance computing (HPC) applications are increasingly shifting from CPU-based implementations to graphics processing units (GPUs). This shift is motivated both by the high arithmetic throughput and by the favorable energy efficiency of GPUs, which typically provide substantially more floating-point operations per unit power than conventional CPUs. Further, the reduction in execution time enables classes of applications that require not just a single large simulation, but a large number of repeated evaluations -- for example simulation-based inference \cite{cramer_2020}. In addition, recent software frameworks such as \textsc{jax} make it possible to combine accelerator-based performance with just-in-time compilation, automatic differentiation, and a high-level programming model, which is particularly attractive for modern scientific applications \cite{jax}.

\begin{figure}
    \includegraphics[width=\columnwidth]{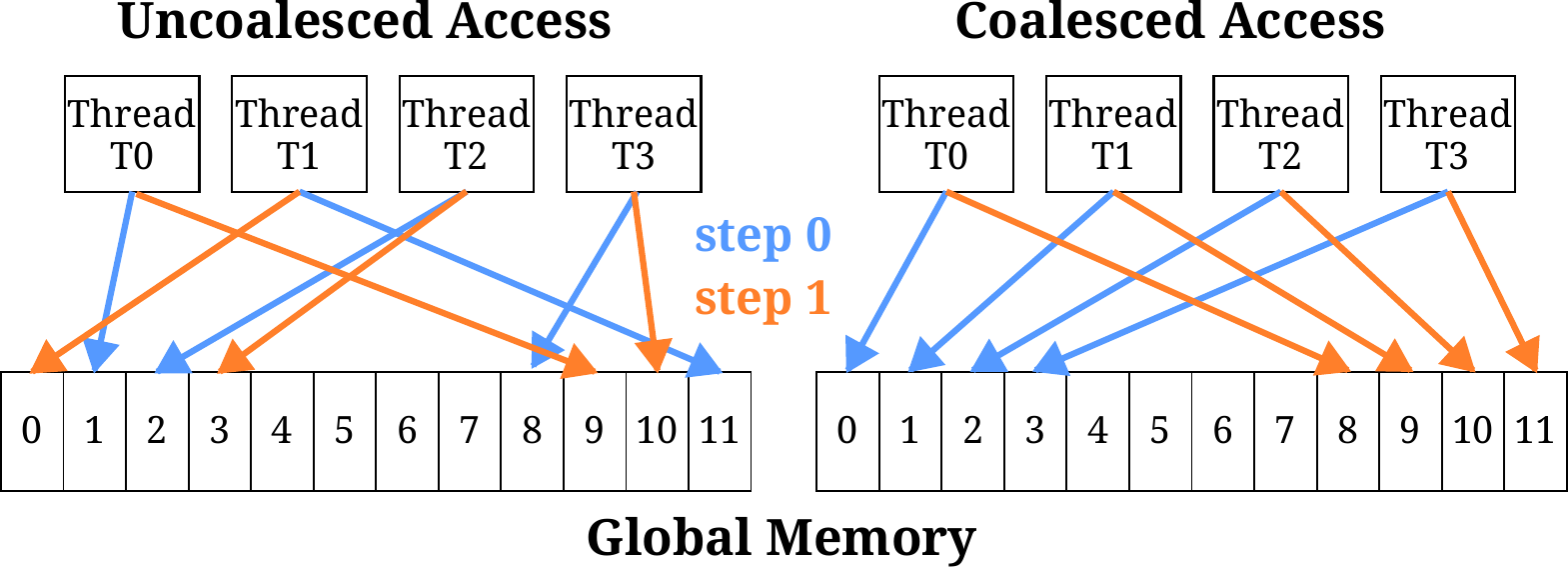}
    \caption{Illustration of uncoalesced versus coalesced memory access. Coalesced memory access patterns achieve significantly better performance on GPU than uncoalesced access.}
    \label{fig:coalescence}
\end{figure}

However, GPUs differ fundamentally from CPUs in how performance is achieved. GPUs follow a throughput-oriented parallel execution model with large numbers of lightweight threads, which differs significantly from the latency-optimized design of CPUs~\cite{cuda_guide}. As a consequence, many algorithms that are state-of-the-art on CPUs perform poorly when transferred directly to GPUs without redesign. Efficient GPU implementations typically require minimizing host-device communication, reducing global memory traffic, limiting thread divergence, and maximizing memory coalescence. In particular, coalesced memory access (illustrated in Figure \ref{fig:coalescence}) is a primary performance consideration, as memory transactions are shared across threads within a warp~\cite{cuda_best_practices}. Similarly, divergent control flow within a warp can significantly degrade performance, since threads executing different branches must be serialized~\cite{cuda_guide}. In practice, these constraints favor algorithms with regular control flow, predictable memory access patterns, and a limited number of synchronization points.

\subsection{Related Work}

Tree-based data structures are a particularly important example of this challenge. On CPUs, trees are a standard tool for reducing the complexity of spatial search and interaction problems~\cite{samet_2006,berg_2008}. They are used in nearest-neighbour search~\cite{bentley_1975}, friends-of-friends clustering, $N$-body methods~\cite{barnes_hut_1986}, multipole schemes~\cite{greengard_rokhlin_1987}, and many related algorithms. Yet on GPUs, tree methods are often much less competitive than their asymptotic complexity would suggest. Tree construction is frequently expensive, traversal tends to induce thread divergence, and the associated memory access patterns are often highly irregular~\cite{karras_2012,lauterbach_2009,zhou_2008}. While iterative traversal schemes can reduce some forms of control-flow divergence \cite{jakob2021optimizing}, they generally do not eliminate divergence in the number of traversal steps taken by different threads.  Further, conventional tree layouts make it difficult for neighbouring threads to read memory collaboratively, so that even moderate divergence in traversal may quickly destroy memory coalescence.

In particular, nearest neighbour search has been studied extensively, and a wide range of algorithmic approaches have been proposed. Classical exact methods are typically based on spatial tree structures such as KD-trees~\cite{bentley_1975} or ball trees, which enable logarithmic query complexity in low dimensions. Variants of dual-tree traversal further improve efficiency by processing interactions between groups of nodes jointly~\cite{gray_2000,curtin_2013}. Closely related approaches based on uniform grids or cell lists are widely used in particle simulations, where the domain is decomposed into regular bins to enable efficient neighbour queries with predictable memory access patterns~\cite{hockney_1988,allen_2017}. On modern hardware, particularly GPUs, brute-force approaches based on dense distance evaluations have become increasingly competitive due to their regular memory access patterns and high arithmetic intensity~\cite{garcia_2008,faiss_2021,faiss_2025}. In addition, a large body of work has focused on approximate nearest neighbour (ANN) methods, including hashing-based techniques and product quantization~\cite{jegou_2011}, as well as graph-based approaches such as nearest-neighbour graphs and navigable small-world structures~\cite{malkov_2020}. These methods often achieve significantly improved query times at the cost of approximation error. Finally, a notable recent advancement for exact neighbour search on GPUs is \textsc{clover}~\cite{kamel_2025}, a spatio-graph-based method that constructs an index of random Voronoi partitions to prune the search space while maintaining high hardware utilization, outperforming prior tree methods by an order of magnitude for some setups.

\subsection{Contributions}

In this work we propose a novel tree framework designed specifically to address the constraints of GPU architectures. The presented hierarchy is based on Morton, or $z$-order, sorting and can be constructed efficiently in a bottom-up fashion. Rather than producing a deeply nested binary tree with irregular traversal depth, the construction yields a hierarchy of tree-planes with fixed and small depth. This makes tree walks highly predictable and allows them to be implemented through a small number of kernel launches. In addition, the hierarchy is organized such that the children of a node are stored contiguously and may be accessed with fully coalesced memory reads. Combined with a dual tree walk formulation, this allows interactions between groups of nodes to be processed collaboratively, reducing redundant memory access compared to more conventional traversal schemes.

We demonstrate the benefits of the framework with two algorithms, $k$-nearest neighbour search and friends-of-friends (FoF) clustering. For both cases, we find significant performance improvements over the closest competiting libraries, reaching more than an order of magnitude improvement for sufficiently large problems. The presented framework is not specific to these two use cases. The same tree representation and traversal strategy can be extended naturally to a range of other tree-based algorithms, including density-based clustering methods such as DBSCAN, fast multipole methods, and correlation function estimation.  

Our main optimization target is for low dimensions ($d \sim 3$), and large point counts $N \gg 10^6$ -- a regime that is highly relevant for many HPC simulation codes. We are less concerned with very high-dimensional settings or with small problem sizes, although we will show that the presented methods remain competitive outside of the primary target regime as well. Here, we only consider a Euclidean distance measure, but including other distance measures in the future would be viable.

The remainder of this paper is structured as follows. In Section~\ref{sec:ztrees} we introduce the $z$-order based tree construction. We then describe the nearest neighbour algorithm and evaluate its performance in Section \ref{sec:knn} and the FoF algorithm in Section \ref{sec:fof}. Finally, we conclude in Section \ref{sec:conclusions}.

The publication is accompanied by an open-source implementation of the presented algorithms named '$\textsc{jz-tree}$' (short for \textsc{jax} z-order tree) that is available on GitHub\footnote{https://github.com/jstuecker/jztree/} and PyPI\footnote{https://pypi.org/project/jztree} with additional documentation and usage examples under\footnote{https://jstuecker.github.io/jztree/}.

\section{Z-order trees} \label{sec:ztrees}
We construct a tree in two steps: (1) A sort of the input position array in Morton / z-order \cite{Morton1966} and (2) A search for splitting points on the position array to summarize points (and nodes) into coarser nodes. Both steps involve only GPU friendly operations on flat arrays so that the tree construction is significantly faster on GPUs than widely used top-down construction methods of KDTrees.

We note that Peano-Hilbert (PH) order is often considered superior to z-order in terms of spatial locality~\cite{sagan_1994, bader_2013}. However, we prefer z-order here due to its simplicity and flexibility. In particular, z-order can be defined directly for all floating-point coordinates without requiring a predefined domain or refinement level. In contrast, PH order is typically constructed on discretized grids and becomes more complex to generalize across dimensions or to arbitrary floating-point data.

\subsection{Z-order sort}

The most common and fastest approach to sorting position vectors in z-order is to sort by an integer key obtained by interleaving the bits of the coordinate components (Morton encoding \cite{Morton1966, Samet2006}). However, for floating-point positions, defining such a key requires restricting the domain and truncating precision. An alternative approach is to define a custom comparison operator that directly compares full position vectors and to use a sorting algorithm that supports custom comparators, such as mergesort~\cite{connor_2010}. Here, we adopt this approach, as it provides maximum generality -- allowing the construction of tree structures at full floating-point precision. As we show later, the associated performance overhead is negligible, since sorting is not a bottleneck in the presented algorithms.

To define this comparison operator, let us assume that we have a function $\text{msb}_{\text{fixed}}$ available that extracts the most significant differing bit of two positive fixed-point numbers (normalized to exponent $0$). For example, for two numbers $a$ and $b$
\[
\begin{array}{r@{\;}l}
a =  \overbrace{...1010}^{\text{common}}&\!1001.0101.. \\
b =  ...1010&\!0101.1011... \\
     &\!\uparrow \\
     &\text{\hspace{-60pt}most significant differing bit at $2^3$}
\end{array}
\]
the most significant differing bit is the first bit at which the two representations differ after their common prefix. We label bits by the power of two that they represent so that $\text{msb}_{\text{fixed}}$ would return $3$ in the example above. Such a function can easily be implemented by counting leading zeros on a bitwise exclusive or of $a$ and $b$. Given the function $\text{msb}_{\text{fixed}}$ we may define a more general function $\text{msb}$ that acts on floating point numbers to extract the most significant bit that would differ if they were normalized as fixed point numbers with exponent 0. Given a separation into sign, exponent and mantissa:
\begin{align}
        a &= \sign(a)  \cdot  2^{\ex(a)} \cdot \mant(a)
\end{align}
where $\sign \in \{-1, 1\}$, $\ex \in \mathbb{N}$ and $\mant \in [1, 2)$, we may write
\begin{align}
        \text{msb}(a, b) &= \begin{cases}
                \text{EMAX} & \text{if  }  \sign(a) \neq \sign(b) \\
                \max(\ex(a), \ex(b)) & \text{else if  }  \ex(a) \neq \ex(b) \\
                \ex(a) + \text{msb}_{\text{fixed}}(\mant(a), \mant(b)) & \text{else}
        \end{cases}
\end{align}
where $\text{EMAX}$ is one larger than the maximum exponent value (e.g., $128$ for float32 and $1024$ for float64 in IEEE 754 standard \cite{ieee754}). The function $\text{msb}$ can be implemented efficiently using bitwise operations (bit shifts and leading-zero counting), with special care required to handle subnormal numbers, where the mantissa representation differs from normalized values.


We can then define the z-order comparison operator of two vectors $\arr{p}, \arr{q} \in \mathbb{R}^d$ as a comparison along the dimension with the largest most significant bit difference:
\begin{align}
        k &= \text{argmax}_i \text{msb}(p_i, q_i) \\
        \arr{p} \prec_{z} \arr{q} &\Leftrightarrow p_k \leq q_k
\end{align}
where $\text{argmax}$ selects the first occurrence of the maximum -- so that differences in earlier coordinates are more significant than those in later coordinates. 

\begin{figure}[!t]
\centering
\includegraphics[width=\columnwidth]{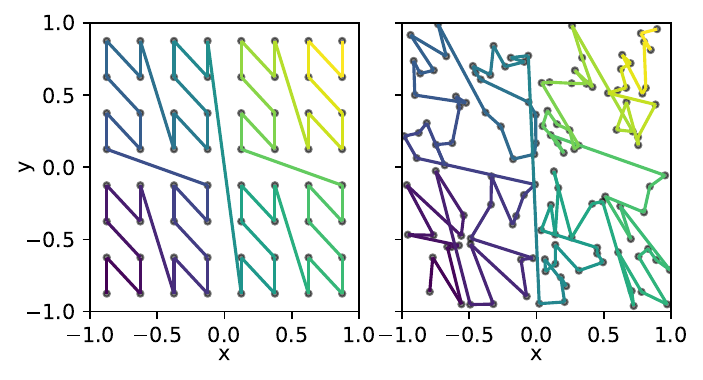}
\caption{The z-order curve for a regular grid (left) and for randomly distributed points (right). Colour encodes the linear index in the sorted array.}
\label{fig:zorder}
\end{figure}

In Figure \ref{fig:zorder} we show two examples of the z-sorted points on a regular grid (left) and for a uniform random distribution (right). For a regular grid, the traversal follows the characteristic Z-shaped (Morton) pattern \cite{Morton1966}.

In \textsc{jz-tree} the z-order sort is implemented via a library call to the mergesort routine of the \textsc{CUB} library\cite{CUB}. For multi-GPU execution, we employ a sampling-based partitioning approach\cite{Shi_1992}. In a first step, a subset of $N_{\mathrm{samp}} \sim 1000$ points is sampled on each GPU, collected, sorted on a single device and broadcasted. Based on the sorted samples, a set of splitters is chosen such that the sampled points are evenly partitioned. Subsequently, all points are redistributed across GPUs according to these splitters and sorted locally, resulting in a globally consistent z-order.

Assuming no duplicate keys, the expected relative imbalance scales as $O(\sqrt{\log(N_{\mathrm{GPU}}) / N_{\mathrm{samp}}})$~\cite{frazer_1970,sanders_2004,Mitzenmacher_Upfal_2005}, leading to imbalances well below $10\%$ in all scenarios that we consider here. After sorting, domain boundaries can be adjusted for perfect load balance.

\subsection{Nodes}

We define a node with center $\arr{c}$ and Morton level $\arr{lvl}$ as the set of all points whose hypothetically interleaved binary representations share the leading bits up to $\arr{lvl}$ with $\arr{c}$. Such a node corresponds to a contiguous segment in z-order.

Given two points $\arr{p}, \arr{q} \in \mathbb{R}^d$, let $k$ denote the dimension in which they differ most significantly in the Morton sense. The corresponding Morton level is
\begin{align}
        \mathrm{lvl}(\arr{p}, \arr{q}) &= (\text{msb}(p_k, q_k) + 1) \cdot d - k.
\end{align}
where the offset by one accounts for the fact that the common interval is larger than the position of the highest differing bit.
We may further define per-dimension extent levels
\begin{align}
        l_i &= \left\lfloor \frac{\mathrm{lvl}}{d} \right\rfloor + 
        \begin{cases}
                1 & \text{if } i \ge (\mathrm{lvl} \bmod d), \\
                0 & \text{otherwise}.
        \end{cases}
\end{align}
so that $L_i = 2^{l_i}$ corresponds to the spatial extent of the node in the $i$th component and $2^\mathrm{lvl}$ corresponds to the volume of a node. Extent levels may differ at most by one across dimensions, so that nodes can be rectangular with axis ratios of at most two.




\subsection{Plane based tree-hierarchy}

\begin{figure}
        \begin{tikzpicture}[
        every node/.style={font=\scriptsize},
        x=0.72cm,
        y=0.42cm
        ]

        \pgfmathsetmacro{\lspace}{1.8}
        \pgfmathsetmacro{\wa}{1.4}
        \pgfmathsetmacro{\wb}{2.0}

        \node[anchor=west] at (-2., 0) {$\arr{x}$};
        \node[anchor=west] at (-2., -1) {$\arr{lvl}$};
        \node[anchor=west] at (-2., -2) {$\arr{n}$};
        \node[anchor=west] at (-2., -2 - \lspace*1.) {$\arr{spl}^{(0)}$ ($n>2$)};
        \node[anchor=west] at (-2., -2 - \lspace*2.) {$\arr{spl}^{(1)}$ ($n>4$)};

        \foreach \p in {0, 1, 2, 3, 4, 5, 6, 7, 8} \draw[opacity=0.3] (\p+0.5, 0.25) -- (\p + 0.5, -2.0);

        \foreach \i/\j in {0/0, 1/1, 2/3, 3/5, 4/6, 5/8} {
                \draw[opacity=0.3] (\wa*\i + 1.,-2-\lspace*1.) -- (0.5 + \j, -2.0);
        }
        \foreach \i in {0, 1, 2, 3, 4} {
                \draw[opacity=0.3] (\wa*\i + 1.,-2-\lspace*1.) .. controls (\wa*\i + 1. + 0.5*\wa,-2-\lspace*1.5) .. (\wa*\i + 1. + \wa,-2-\lspace*1.);
                \node[anchor=center] at (\wa*\i + 1. + \wa*0.5,-2-\lspace*1.5) {Leaf \i};
        }
        \foreach \i/\j in {0/0, 1/2, 2/4, 3/5} \draw[opacity=0.3] (1.5 + \wb*\i,-2-\lspace*2.) -- (1. + \wa*\j,-2-\lspace*1.);
        \foreach \i in {0, 1, 2} {
                \draw[opacity=0.3] (\wb*\i + 1.5,-2-\lspace*2.) .. controls (\wb*\i + 1.5 + 0.5*\wb,-2-\lspace*2.5) .. (\wb*\i + 1.5 + \wb,-2-\lspace*2.);
                \node[anchor=center] at (\wb*\i + 1.5 + \wb*0.5,-2-\lspace*2.5) {Node \i};
        }

        \foreach \p/\v in {
        0/{(-\infty)}, 1/{1.6}, 2/{3.1}, 3/{3.3}, 4/{4.6},
        5/{5.6}, 6/{6.8}, 7/{9.4}, 8/{9.7}, 9/{(\infty)}
        } \node at (\p,0) {$\v$};

        \foreach \p/\v in {
        0.5/{(129)}, 1.5/{2}, 2.5/{-1}, 3.5/{3}, 4.5/{1},
        5.5/{2}, 6.5/{4}, 7.5/{0}, 8.5/{(129)}
        } \node at (\p,-1) {$\v$};

        \foreach \p/\v in {
        0.5/{(8)}, 1.5/{3}, 2.5/{2}, 3.5/{6}, 4.5/{2},
        5.5/{3}, 6.5/{8}, 7.5/{2}, 8.5/{(8)}
        } \node at (\p,-2) {$\v$};

        \foreach \i/\j in {0/0, 1/1, 2/3, 3/5, 4/6, 5/8} \node at (\wa*\i + 1.,-2-\lspace*1.) {$\j$};
        \foreach \i/\j in {0/0, 1/2, 2/4, 3/5} \node at (\wb*\i + 1.5,-2-\lspace*2.) {$\j$};

        \end{tikzpicture}
        \caption{Illustration of the steps involved in building a split based tree structure. By choosing splitting points that have $n > N_{\mathrm{max}}$ we obtain tree-planes where all nodes have $n \leq N_{\mathrm{max}}$ points.}
        \label{fig:tree_construction}
\end{figure}
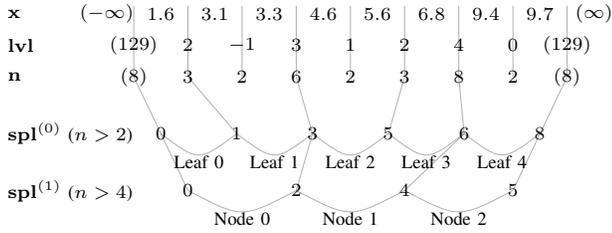

As a first step to constructing a tree hierarchy we calculate
\begin{align}
        \mathrm{lvl}_i = \mathrm{lvl}(\arr{x}_{i-1}, \arr{x}_{i})
\end{align}
for each consecutive pair of points $i-1$ and $i$ in the sorted array. To simplify later calculations, we assume the existence of an additional vector with $-\infty$ components at index $-1$ and $+\infty$ components at index $N$, so that we obtain $N+1$ values for $N$ points. It is useful to interpret these level values as being associated with the gaps between consecutive points (see Figure~\ref{fig:tree_construction}).

As a next step, we determine the range of points contained in the smallest node that includes points $i-1$ and $i$. To this end, we perform a binary search to the left to find the smallest index $l_b$ that would be part of such a node
\[
\mathrm{lvl}(\arr{x}_{l_b}, \arr{x}_i) \le \mathrm{lvl}_i,
\]
and a binary search to the right to find the smallest index $r_b$ that would be outside
\[
\mathrm{lvl}(\arr{x}_{i-1}, \arr{x}_{r_b}) > \mathrm{lvl}_i.
\]
If no such indices exist, we set $l_b = 0$ and $r_b = N$. The number of points contained in the node is then $n = r_b - l_b$.

The information contained within $l_b$ and $r_b$ is in principle sufficient to define a full binary tree, where the parent of each node is given by $l_b$ or $r_b$ depending on which one has the lower level $\mathrm{argmin}(\mathrm{lvl}_{l_b}, \mathrm{lvl}_{r_b})$. However, walking such a binary tree on GPU architectures would lead to poor memory coalescence, since different threads may access very different locations in memory. We therefore choose a different approach here, where we allow nodes to have a variable number of children, but keep the depth of the resulting tree fixed.

We define a \emph{tree-plane} as a set of nodes that partitions the points $\arr{x}$, such that each point belongs to exactly one node (while empty regions of space may remain uncovered). A tree-plane may be parameterized through a set of $N_{\mathrm{nodes}+1}$ splitting points $\arr{spl}$ in the z-order index space so that a node $i$ contains all points in the range $[\mathrm{spl}_i, \mathrm{spl}_{i+1})$. Recall that $n_i$ is the number of points that would need to be included in a node that contains points $i-1$ and $i$. We construct a tree-plane by selecting all separation points with $n > N_{\mathrm{max}}$. Intuitively, each tree-plane partitions the points into the largest possible Morton cells subject to the constraint that each cell contains at most $N_{\mathrm{max}}$ points.
In Figure \ref{fig:tree_construction} we illustrate the splitting points $\arr{spl}^{(0)}$ of leaf nodes created this way with $N_{\mathrm{max}}^{(0)} = 2$ which we refer to as the '0th plane'.

\begin{figure}[!t]
        \centering
        \includegraphics[width=\columnwidth]{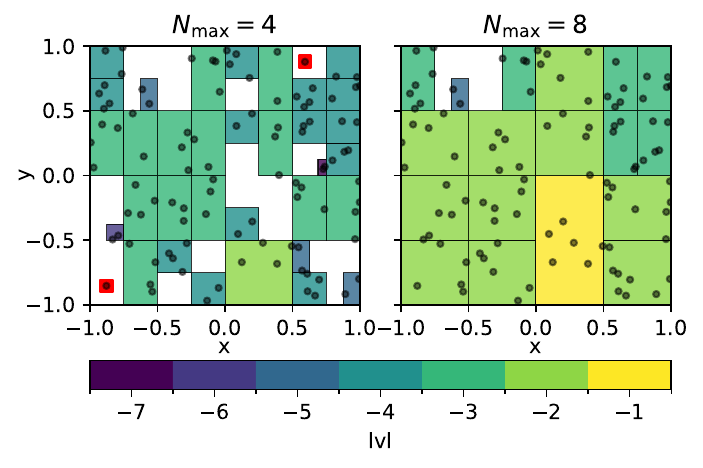}
        \caption{Nodes that are obtained by selecting the largest Morton nodes that hold at most $N_{\mathrm{max}}=4$ (left) or $N_{\mathrm{max}}=8$ (right) points. These nodes only fill the space that contains points and some nodes may only contain a single points leading to zero extent (marked with red squares).}
        \label{fig:tree_planes}
\end{figure}

We may construct coarser tree-planes iteratively by applying the same procedure to the splitting points of the previous plane. That is, we retain only those splits between nodes of plane $p$ for which $n$ exceeds $N_{\mathrm{max}}^{(p+1)}$.
In Figure \ref{fig:tree_planes} we show an example of two tree-planes that are obtained with $N_{\mathrm{max}}^{(0)}=4$ and $N_{\mathrm{max}}^{(1)}=8$ for a uniform random distribution on a two-dimensional domain in the range $[-1, 1]$ with $N=100$ points. Note a few properties that are different between this tree-plane hierarchy and conventional space-filling binary trees:
\begin{itemize}
        \item Space that doesn't contain points may or may not be part of a tree-node.
        \item Some nodes may only contain a single point and have zero extent.
        \item Nodes on the coarser plane may contain a flexible number of nodes of the finer plane.
        \item Some nodes on the coarser plane may have themselves as their own only child on the finer plane.
        \item Different children may have different extent.
        \item The tree has the same fixed depth everywhere.
\end{itemize}

In practice we build a tree-plane hierarchy by choosing $N_{\mathrm{max}}^{(0)}$ to define leaf nodes for the finest level of the tree and then successively increasing by factor $c$ per plane level:
\begin{align}
        N_{\mathrm{max}}^{(p)} &= N_{\mathrm{max}}^{(0)} c^p
\end{align}
with defaults $N_{\mathrm{max}}^{(0)} = 48$ and $c = 8$ which we find to be good choices performance wise. If we wanted to end up with a single root node, we could keep coarsening until $N_{\mathrm{max}}^{(p)} \geq N$ at which point we'd be guaranteed to have a single node that covers all points. However, on GPUs it is preferable to have a coarsest level that has already a notable number of nodes so that most streaming multiprocessors have work to do from the beginning. We define a target number $N_{\mathrm{target}}$ (typically of order 1000) that we aim to obtain at the coarsest level. We may get a rough estimate of the number of nodes that might be contained on a tree-plane with $N_{\mathrm{max}}$ based on the heuristic that typical nodes should contain at least $N_{\mathrm{max}}/2$ points, since otherwise they might be merged with one of their neighbours:
\begin{align}
        N_{\mathrm{nodes}} \lesssim \frac{N}{N_{\mathrm{max}}/2}
\end{align}
This typically overestimates the number of nodes, but it is not a strict upper bound, since in z-order a single high-$n$ node may block multiple low-$n$ nodes from merging. We stop coarsening when the estimated number of nodes is smaller than $N_{\mathrm{target}}$.

In \textsc{jz-tree} the distributed tree-construction is implemented in four steps: (1) We first adjust the domain to ensure that no node of the coarsest tree-plane crosses domain boundaries.
This is achieved by determining how far a node at a given Morton level would extend across the domain boundary. The domain then needs to be adjusted to the starting or end point of the largest node that contains $\leq N_{\mathrm{max}}^{(\mathrm{last})}$ points. The subsequent tree construction can be treated fully locally from this point on. (2) We extract leaf splitting points $N_{\mathrm{max}}^{(0)}$ by checking where splitting points exceed $N_{\mathrm{max}}^{(0)}$ through a range search of $N_{\mathrm{max}}^{(0)}$ to the left and right of each splitting point. This step is optional, but tends to be faster for the leaf level than a binary search. (3) We then determine $n$  for each leaf splitting point through the earlier described binary search and (4) Extract splitting points for the full hierarchy.

\subsection{Regularization}

For many problem setups (e.g. uniform random distributions or particle distributions from cosmological simulations), the described tree structure is sufficient. However, for distributions that contain a small number of points far from the bulk -- e.g. multivariate Gaussian distributions -- summarizing nodes solely based on the number of contained points may produce a small number of nodes with very large extent. 
This is problematic for nearest-neighbour search, where at least a region of size comparable to the node must be explored, which in the worst case can include almost all points.


To improve performance in such scenarios, we introduce a simple regularization criterion. For each tree-plane $p$, we define a global maximum level $\mathrm{lvl}_{\mathrm{max}}^{(p)}$ and retain all splitting points whose level satisfies $\mathrm{lvl} > \mathrm{lvl}_{\mathrm{max}}^{(p)}$. Intuitively, this prevents the formation of excessively large nodes in low-density regions by enforcing a global upper bound on node size.
We define this maximum level so that the volume $V_{\mathrm{max}}^{(p)} = 2^{\mathrm{lvl}_{\mathrm{max}}^{(p)}}$ of nodes never exceeds
\begin{align}
    V_{\mathrm{max}}^{(p)} = 2^{\mathrm{lvl}_{\mathrm{max}}^{(p)}} = f_{\mathrm{max}} V_{90\%}^{(p)}
\end{align}
where we typically choose $f_{\mathrm{max}} \sim 50$. Here, $V_{90\%}^{(p)}$ denotes the point number weighted average volume of nodes on plane $p$, computed over the subset of smallest nodes that together contain $90\%$ of all points.
This excludes a small number of very large nodes that may otherwise dominate the average.

For the scenarios considered in this work, this simple regularization scheme is sufficient. However, we leave the possibility of incorporating more sophisticated techniques in \textsc{jz-tree} for future work.

\subsection{Multiple point types}
Some algorithms require treating multiple point types separately in the tree. For example, in nearest-neighbour search, one may wish to query the tree using a set of query points $\arr{x}_{\mathrm{query}}$ distinct from the source points $\arr{x}$. 

The most common approach is to construct a separate tree for the query points and to treat query and source trees explicitly during the dual tree walk \cite{gray_2000, ram_2009, curtin_2013}. However, this increases implementation complexity and may lead to suboptimal refinement, as the query tree is constructed independently of the source distribution.

Instead, we construct a single tree jointly over all point types. This is achieved by concatenating the positions of all types into a single array prior to the $z$-order sort. During tree construction, we track the point counts of each type separately for every candidate node. Splitting points are then chosen such that the maximum count over all types does not exceed $N_{\mathrm{max}}^{(p)}$ for any node on tree-plane $p$.

After construction, points are separated again into type-specific arrays while preserving $z$-order. Only the leaf-level splits $\arr{spl}^{(0)}$ are defined separately for each type. This enables coalesced memory access within each species while maintaining a shared tree structure that adapts to all point distributions and keeps the tree traversal simple.

\subsection{\textsc{jax} implementation details}

Most computationally intensive parts of our implementation are realized as \textsc{CUDA} kernels, invoked via the foreign function interface (FFI) of \textsc{jax}. To maintain compatibility with \textsc{jax}'s just-in-time (JIT) compilation, all memory allocations must have statically known sizes at jit-compile time. 

Since the number of nodes per tree-plane is data-dependent, we allocate one contiguous buffer for each node property (e.g. splitting points, particle counts, node centers, node levels) and store all tree-planes within this buffer using data-dependent offsets.

The required allocation size is estimated as
\begin{align}
    \text{allocation\_size} = \text{alloc\_fac\_nodes} \cdot \frac{N}{N^{(0)}_{\mathrm{max}}/2}, \label{eqn:alloc_nodes}
\end{align}
where typically $\text{alloc\_fac\_nodes} \sim 1\text{--}2$ is sufficient in practice. If the allocated size is insufficient, a runtime error is raised indicating the required increase.

\section{Nearest neighbour search} \label{sec:knn}
We describe how to implement a $k$-nearest neighbour search based on the plane-based tree hierarchy that we have described in Section \ref{sec:ztrees}.  The neighbour search happens conceptually in two steps: (1) A dual tree walk on the tree hierarchy to determine per leaf an interaction list of other leaves that need to be checked to guarantee that all candidate neighbours required for an exact $k$-nearest neighbour search are considered. (2) A neighbour search that traverses the leaf-leaf interaction list collaboratively among points in the same leaf.

\subsection{Interaction lists}
We parameterize an interaction list as a tuple $\arr{ilist} = (\arr{isrc}, \arr{ispl})$ of two arrays: a set of source indices $\arr{isrc}$ and a set of splitting points $\arr{ispl}$. The interaction list is sorted by receiving nodes so that a receiving node $i$ has to interact with the $\arr{isrc}$ indices in the range from  $\mathrm{ispl}_i$ up to $\mathrm{ispl}_{i+1}-1$.

A dense interaction list where every node out of $N_{\mathrm{nodes}}$ interacts with every other node can be initialized as
\begin{align}
        \mathrm{ispl}_i &= N_{\mathrm{nodes}} \cdot i \\
        \mathrm{isrc}_j &= j \text{ mod } N_{\mathrm{nodes}}
\end{align}
for $i \in [0...N_{\mathrm{nodes}}]$ and $j \in [0...N_{\mathrm{nodes}}^2]$.

\subsection{Dual Tree Walk}

\begin{algorithm}[t]
\caption{Dual tree walk for nearest neighbour search}
\label{alg:knn_dual_walk}
\begin{algorithmic}[1]
\REQUIRE For each plane $p=0,\dots,P-1$: node splits $\arr{spl}^{(p)}$, source point count $\arr{n}^{(p)}$, node centers $\arr{c}^{(p)}$, and Morton levels $\arr{lvl}^{(p)}$. 
Source positions $\arr{x}$ and leaf splits $\arr{spl}^{(0)}$. Query positions $\arr{x}_{\mathrm{q}}$ and splits $\arr{spl}^{(0)}_{\mathrm{q}}$.
\STATE $\arr{spl}^{(P)} \gets \textsc{Range}(0, N_{\mathrm{topnodes}}, \text{NGR})$
\STATE $\arr{ilist} \gets \textsc{DenseInteractionList}(\lceil N_{\mathrm{topnodes}} / \text{NGR} \rceil)$
\FOR{$p=P-1$ down to $0$}
    \STATE $\arr{ilist} \gets \textsc{NodeToNode}(\arr{ilist}, \arr{spl}^{(p+1)}, \arr{n}^{(p)}, \arr{c}^{(p)}, \arr{lvl}^{(p)})$
\ENDFOR

\RETURN $\textsc{LeafToLeaf}(\arr{ilist}, \arr{spl}^{(0)}, \arr{x}, \arr{spl}^{(0)}_{\mathrm{q}}, \arr{x}_{\mathrm{q}})$
\end{algorithmic}
\end{algorithm}

We sketch the necessary steps for the dual tree walk in Algorithm~\ref{alg:knn_dual_walk}. As a first step, we group the top level nodes into 'pseudo' super nodes where $\text{NGR}$ denotes the grouping size.
This grouping is necessary because an entry in the interaction list represents interactions between all children of the receiving node and all children of the source node. Grouping ensures that this assumption remains valid at the top level.
A dense interaction list is then initialized on these super nodes so that effectively every top-node will interact with every other top-node. The precise value of NGR is not critical and we typically choose $\text{NGR}=32$. Subsequently, we evaluate a node-node interaction function on every plane to move the interaction list from plane $p+1$ to $p$ and finally evaluate the leaf-leaf interaction list.

Given two nodes with centers $\arr{c}_1$ and $\arr{c}_2$ and per-dimension extent vectors $\arr{L}_1$ and $\arr{L}_2$ (that may be calculated from the Morton level $\arr{lvl}$), we can define a lower distance $d_{\mathrm{low}}$ and an upper distance $d_{\mathrm{up}}$ as
\begin{align}
        \mathrm{s}_{\pm, i}(\arr{a}, \arr{b}) &= \max(|a_i| \pm b_i, 0) \\
        d_{\mathrm{low}} &= \left\lVert \arr{s}_{-}\left(\arr{c}_1 - \arr{c}_2, \frac{\arr{L}_1 + \arr{L}_2}{2} \right) \right\rVert \\
        d_{\mathrm{up}} &= \left\lVert \arr{s}_{+}\left(\arr{c}_1 - \arr{c}_2, \frac{\arr{L}_1 + \arr{L}_2}{2} \right) \right\rVert
\end{align}
It is guaranteed that every point in node 1 includes all points from node 2 at a radius $R_{\mathrm{max}} \geq d_{\mathrm{up}}$. Further, it is guaranteed that no point of node 2 lies within a radius smaller than $d_{\mathrm{low}}$ from any point in node 1. We can therefore use $d_{\mathrm{up}}$ to find guaranteed upper bounds for the radius in which neighbours need to be checked and $d_{\mathrm{low}}$ for efficient pruning of interactions.

\begin{algorithm}[t]
\caption{Node to Node interaction function.}
\label{alg:knn_n2n}
\begin{algorithmic}[1]
\REQUIRE Interaction list $\arr{ilist}$, node splits $\arr{spl}^{(p+1)}$, point count $\arr{n}^{(p)}$, centers $\arr{c}^{(p)}$, and Morton levels $\arr{lvl}^{(p)}$.
\STATE $\arr{R}_{\mathrm{max}} \gets \textsc{FindRmax}(\arr{ilist}, \arr{spl}^{(p+1)}, \arr{n}^{(p)}, \arr{c}^{(p)}, \arr{lvl}^{(p)})$
\STATE $\arr{icount} \gets \textsc{Count}(\arr{ilist}, \arr{spl}^{(p+1)}, \arr{c}^{(p)}, \arr{lvl}^{(p)}, \arr{R}_{\mathrm{max}})$
\STATE $\arr{ispl} \gets \textsc{CumulativeSumPrep0}(\arr{icount})$
\STATE $\arr{isrc} \gets \textsc{Insert}(\arr{ilist}, \arr{spl}^{(p+1)}, \arr{c}^{(p)}, \arr{lvl}^{(p)}, \arr{R}_{\mathrm{max}}, \arr{ispl})$
\RETURN Interaction list $(\arr{isrc}, \arr{ispl})$
\end{algorithmic}
\end{algorithm}

The node-to-node interaction function is sketched in Algorithm~\ref{alg:knn_n2n}. It works in four steps, each of which requires a separate kernel launch: (1) Determine for each node a maximum radius $\arr{R}_{\mathrm{max}}$ that guarantees that it contains the k-th nearest neighbour of all points inside the node. (2) For each node, count the number of nodes for which $d_{\mathrm{low}} \leq R_{\mathrm{max}}$. (3) Calculate the cumulative sum (and prepend 0). (4) Insert the interaction source indices using $\arr{ispl}$ as relative offsets in the array.

Steps (1), (2) and (4) share very similar traversal logic, so we will only discuss step (1) in detail to highlight how the presented data structures can be used to define a CUDA kernel with a good memory access pattern. The prefix sum in step (3) can be implemented through a library call to \textsc{CUB}.

\begin{algorithm}[t]
\caption{Conceptual outline for \textsc{FindRmax} kernel.}
\label{alg:knn_rmax}
\begin{algorithmic}[1]
\REQUIRE $\text{par\_i}$, $(\arr{isrc}, \arr{ispl}) = \arr{ilist}$, $\arr{spl}^{(p+1)}$, $\arr{n}^{(p)}$, $\arr{c}^{(p)}$, $\arr{lvl}^{(p)}$
\FOR{group step $\text{node\_i} = \arr{spl}[\text{par\_i}]$ up to $\arr{spl}[\text{par\_i} + 1] - 1$}
    \STATE read node data of node\_i into registers
    \STATE $H \gets$ empty neighbour heap with counts
    \FOR{$\text{int} = \arr{ispl[par\_i]}$ up to $\arr{ispl[par\_i + 1] - 1}$}
        \STATE $\text{par\_j} \gets \arr{isrc}[\text{int}]$
        \STATE read node data in range $\arr{spl}[\text{par\_j}] ... \arr{spl}[\text{par\_j} + 1]-1$  into shared memory
        \FOR{$\text{node\_j} = \arr{spl}[\text{par\_j}]$ up to $\arr{spl}[\text{par\_j} + 1] - 1$}
            \STATE $r \gets d_{\mathrm{up}}(\text{node}\_i, \text{node}\_j)$
            \IF{$r < \textsc{RadiusOfCount}(H, k)$}
                \STATE insert $(r, \arr{n}[\text{node}\_j])$ into $H$
            \ENDIF
        \ENDFOR
    \ENDFOR
    \STATE $R_{\mathrm{max}} [\text{node\_i}] \gets \textsc{RadiusOfCount}(H, k)$
\ENDFOR
\end{algorithmic}
\end{algorithm}

The kernel for determining $R_{\mathrm{max}}$ is outlined in Algorithm \ref{alg:knn_rmax}. 
Each thread block is assigned a parent node $\text{par}_i$, determined by the CUDA block index. The outer loop assigns child nodes $\text{node}_i$ of the parent $\text{par}_i$ to individual threads. If the number of children exceeds the number of threads, multiple iterations are required. Subsequently, the interaction list is traversed over source parent nodes par\_j. To minimize global memory access, the data of all child nodes of $\text{par}_j$ is loaded collaboratively into shared memory. Finally, the loop in line 7 iterates over all child nodes (for each thread) to insert the upper node-node distance into the neighbour heap $H$.

The heap data structure $H$ is implemented fully in registers -- following \cite{jakob2021optimizing}. It keeps track of a static number of $\mathrm{N}_r$ distances and counts. $\textsc{RadiusOfCount}$ gives a preliminary estimate of $R_{\mathrm{max}}$ as the smallest radius for which the cumulative count exceeds or equals $k$. If the total count is smaller than $k$, this estimate is set to $\infty$. New entries are inserted into the heap to maintain order, discarding the last element. However, if discarding the last element would lead to the heap holding a total count smaller than $k$, then we instead add the new count to the first element with larger radius. 

The memory access pattern of the \textsc{FindRmax} kernel is ideal for GPU architectures: The global memory accesses in line 2 and line 6 are perfectly coalesced. Further, evaluating interactions between the parent nodes requires only reading each of their children once. This significantly reduces memory access compared to conventional tree walks based on Euler tours where such interactions may be encountered at separate points in time. However, it is worth noting that some threads may be idle if the number of children in par\_i is smaller than the number of threads in the group. E.g. if we choose a coarsening factor of $8$, we'd expect typical nodes to have 8 children which is notably smaller than the minimal number of threads in a group of $32$. In principle, this aspect could be further optimized by assigning multiple threads to the same node and then collaboratively inserting neighbours into a joint heap among those threads. However, we do not attempt this optimization here, because it is only a minor concern for leaf-leaf interactions (where $N_{\rm{max}}^{(0)} \sim 32-64$) which tend to dominate the cost of the neighbour search.

The \textsc{Count} and \textsc{Insert} kernels follow the same structural pattern, but instead of maintaining a heap, they simply count the number of node-node interactions with $d_{\mathrm{low}}(\text{node\_i}, \text{node\_j}) \leq R_{\mathrm{max}}$ and insert the corresponding $\text{node\_j}$ indices into the interaction list.

Finally, the implementation of the \textsc{LeafToLeaf} kernel is again similar to the \textsc{FindRmax} kernel. In this case the outer loop runs over query points (assigning one query point per thread) and the inner loop runs over source points. The neighbour heap structure in this case keeps track of $k_{\mathrm{max}}$ radii and point indices that are written out at the end of each query point iteration. To limit register pressure we choose $k_{\mathrm{max}} \leq 32$ and call the kernel multiple times if $k > k_{\mathrm{max}}$, filtering additionally by a minimum radius $R_{\mathrm{min}}$ (and an equality breaking index offset) that excludes points that were found in previous iterations.

\subsection{Multi-GPU} \label{sec:distr_knn}
Adapting the presented algorithm to multi-GPU scenarios is relatively straightforward. The main idea is that each GPU maintains the local receiving nodes and their corresponding interaction list. Remote source nodes that need to be interacted with are requested once for the evaluation of each plane.

Concretely, the main required additions are as follows: (1) We need to additionally store a tuple of two arrays $\arr{origin} = (\arr{rank}, \arr{idx})$ that saves the origin rank and index for each (unique) source node that appears in the interaction list. (2) When initializing the dense interaction list and super nodes in lines 1-2 of Algorithm~\ref{alg:knn_dual_walk}, $N_{\mathrm{topnodes}}$ includes all (local or remote) top-nodes and $\arr{origin}$ must be initialized appropriately. (3) Before line 4 in Algorithm~\ref{alg:knn_dual_walk} the remote child data $\arr{n}^{(p)}$, $\arr{c}^{(p)}$, $\arr{lvl}^{(p)}$ must be requested for each remote $\arr{origin}$. The corresponding remote splits $\arr{spl}^{(p+1)}$ must be communicated as well. The received data is then rearranged such that $\arr{spl}^{(p+1)}$ correctly indexes contiguous locally available memory. In addition, $\arr{origin}$ is propagated to the child level. (4) After line 4 in Algorithm~\ref{alg:knn_dual_walk} all source indices that appear 0 times in the final interaction list can be removed from $\arr{origin}$. (5) Before line 6 of Algorithm~\ref{alg:knn_dual_walk} we need to do a similar request of leaf splits and source point data.

The strength of this approach is that remote source nodes required for interactions are requested only once, the number of communication points in the algorithm remains small and predictable and the remaining functions remain exactly identical to the single-GPU case. In the scenarios that we have tested, we find that the additionally required remote data is O(10\% - 60\%) of the local receiving node data with a notable dependence on the problem setup and the number of source points per GPU (more data tends to imply better balance). Since it is difficult to foresee all the complications that may arise with more complicated setups and at very large GPU counts, we consider the distributed kNN implementation in \textsc{jz-tree} to be experimental and preliminary.

\subsection{Implementation Details}
We enhance the presented algorithms with an additional component that allows more efficient early pruning in the iteration through interaction lists. For each interaction we additionally store the interaction radius $\arr{r}_{\mathrm{low}}$ -- corresponding to the lower node-node distance of the interaction. For each receiving node we sort $\arr{r}_{\mathrm{low}}$ (after line 4 in Algorithm \ref{alg:knn_n2n}) using a bitonic sort network applied to the corresponding segments. We simply initialize these radii to $0$ at the top-node level. 

This improves the performance for two reasons: (1) Since close-by interactions are encountered earlier, the preliminary estimate of $R_{\mathrm{max}}$ in Algorithm \ref{alg:knn_rmax} is better and more candidate radii can be discarded early (rather than triggering a more expensive insertion into the neighbour heap). (2) It allows to define an early exit after line 5 of Algorithm \ref{alg:knn_rmax} and all other kernels that follow a similar structure: If the maximum current estimate of $R_{\mathrm{max}}$ across all threads is smaller than $\arr{r}_{\mathrm{low}}$, we can discard all remaining interactions. In practice, this prunes on the order of $50\%$ of evaluated interactions.

Finally, we note again that we need to predict allocation sizes at compile time to enable jit-compilation in \textsc{jax}. The main additional allocation that we need to predict here is the size of the interaction list source indices $\arr{isrc}$ (and radii $\arr{r}_{\mathrm{low}}$). Similar to equation~\eqref{eqn:alloc_nodes}, we phrase this allocation relative to the estimated node number:
\begin{align}
    \text{allocation\_size\_ilist} = \text{alloc\_fac\_ilist} \cdot \frac{N}{N^{(0)}_{\mathrm{max}}/2}. \label{eqn:alloc_ilist}
\end{align}
For $d=3$ dimensions, we find that $\text{alloc\_fac\_ilist} \sim 200$ is typically enough, but we note that it is advisable to choose slightly larger values to decrease the chance that the jit-compiled function needs to be aborted due to insufficient available space.

Our primary focus in this article is to optimize performance and memory coalescence to point out a path forward to more GPU friendly tree algorithms. However, it is worth noting that the approach at hand does come at a notable memory cost: With $N^{(0)}_{\mathrm{max}} \sim 48$ and $\mathrm{alloc\_fac\_ilist} \sim 200$ the $\arr{isrc}$ and $\arr{r}_{\mathrm{low}}$ arrays each require the allocation of about $10 \cdot N$ integer / floating point numbers. If only a small number of neighbours $k \lesssim 10$ is requested, this may be the peak contribution to the total required allocation. Further, \textsc{jax}'s memory management system makes it difficult to guarantee that no unnecessary copies of arrays are created. Our implementation in \textsc{jz-tree} is therefore relatively memory-intensive -- something that we aim to improve in future releases.

\subsection{Performance breakdown} \label{sec:knn_performance}
All performance measurements throughout this article are run on the booster nodes of the Leonardo cluster at CINECA \cite{turisini_2024_leonardo}. Each node has a single 32 core Intel Xeon Platinum 8358 processor, four NVIDIA Ampere A100-64 GPUs and 200 Gbps NVIDIA Mellanox HDR InfiniBand connection. Tests with up to 4 GPUs run on a single node, and larger tests run across several nodes (if $N_{\mathrm{GPU}} \geq 4$). For CPU codes we consider tests for a single core and a 32 core setup on a single node.

\begin{figure}
    \includegraphics[width=\columnwidth]{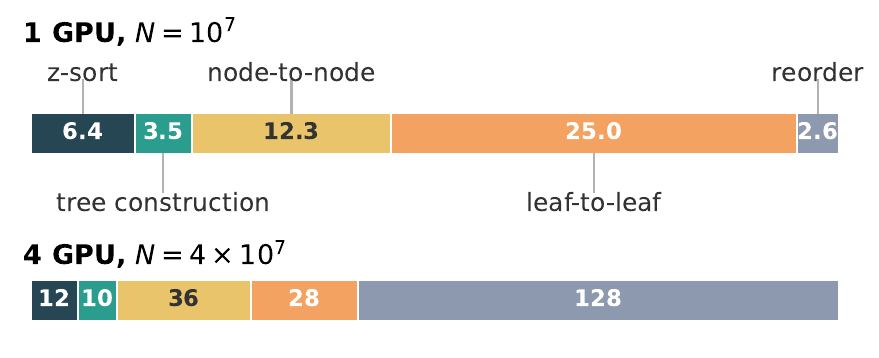}
    \caption{Execution time in ms of different parts of the neighbour search algorithm for returning $k=16$ neighbours in input order for a single GPU setup with $10^7$ points (top) and a 4-GPU setup with $4 \times 10^7$ points. The expensive final reordering step can be avoided in most applications.}
    \label{fig:knnsteps}
\end{figure}

In Figure \ref{fig:knnsteps} we break down the execution time of different steps of a self-neighbour search for a uniform random distribution in three dimensions for a single-GPU and a multi-GPU scenario. The single-GPU case highlights the very low cost of the sorting and the tree construction (about $20\%$ of the total). The most expensive part of the algorithm are the leaf-to-leaf interactions -- comprising approximately $50\%$ of the total execution time. This is expected due to the high computational intensity of this step.

However, for the multi-GPU scenario the costs of several steps increases significantly: The z-sort due to the required exchange of points, the tree construction due to the communication step required for regularization and the node-to-node interactions due to multiple required all-to-all communications and the cost of removing unused nodes from the interaction list. Noteworthily, the leaf-to-leaf interactions only require slightly more time, since they only need a single communication with relatively low volume (thanks to efficient pruning from higher levels). The cumulative effect of these steps is an approximate factor 2 decrease in efficiency. 

However, the most significant increase in execution time is due to the final reordering. This is not too surprising, since bringing the neighbour list into input order requires an extremely high volume communication (recall that these are $k=16$ radii and indices per point). Fortunately, in many applications of neighbour search, it is possible to perform a reduction operation while maintaining the neighbour list in z-order and then only communicate back some small summary statistic per point. We provide a simple interface for this recommended usage pattern in $\textsc{jz-tree}$ and we output points in z-order for further multi-GPU benchmarks, staying representative of such uses cases.

\subsection{Performance comparisons}

\begin{figure}
    \includegraphics[width=\columnwidth]{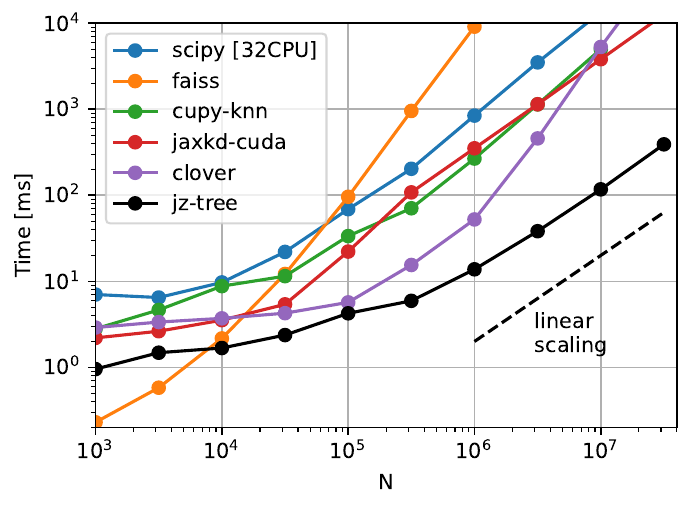}
    \caption{Comparison to other KNN libraries for a uniform random distribution in d=3 dimensions.}
    \label{fig:knnlib}
\end{figure}

We compare the performance of \textsc{jz-tree} for a kNN-search against other publicly available (exact kNN) libraries in Figure \ref{fig:knnlib} for a single GPU setup. The benchmark is to find the $k=30$ nearest neighbours\footnote{In general we use $k=16$ as a baseline in benchmarks, but here we use $k=30$ to allow comparison with the default setup in \textsc{clover}.} for a uniform random distribution of points on the range $[0,1]^d$ in $d=3$ dimensions for $N$ separate source and query points at float32 precision. In each case we include preparation steps (e.g. sorting and tree building) in the performance measurement, so that this represents fairly the total time that is needed to evaluate one set of source points with one set of query points. However, we exclude the jit compilation time that is necessary in \textsc{jax} and \textsc{cupy} implementations. 

The libraries that we compare to are: (1) \textsc{scipy-ckdtree}  -- a CPU based kd-tree library implemented as a \textsc{C++} extension within \textsc{SciPy}, operating on \textsc{NumPy} arrays \cite{scipy}. We include measurements for usage of $1$ and $32$ worker threads. (2) The \textsc{FAISS} library that provides efficient implementations of brute force neighbour search\cite{faiss_2021,faiss_2025}. (3) \textsc{cupy-knn} that implements neighbour search through a one-sided traversal of kd-trees in CUDA kernels\cite{jakob2021optimizing}. (4) Similarly, \textsc{jaxkd-cuda} based on the \textsc{cudaKDTree} library, but offering a convenient jax interface \cite{dodge_jaxkd_2024, dodge_jaxkd_cuda_2024, wald2025stackfree}. (5) \textsc{clover} which traverses a graph based on a random voronoi tessellation\cite{kamel_2025}.

$\textsc{jz-tree}$ outperforms all competitor libraries by a significant margin for nearly all problem sizes (except the brute-force approach of $\textsc{FAISS}$ at very small problem sizes $N \lesssim 10^4$ where the cost of the many kernel launches leads to an irreducible overhead of $\sim 1$ms.) For $N \lesssim 10^6$ \textsc{clover} remains the closest competitor (within about a factor 2), but at larger problem sizes $N \gg 10^6$ $\textsc{clover}$ starts scaling quadratically making it more than an order of magnitude slower at $N \sim 10^7$. The kd-tree based libraries all exhibit the same (close to linear) asymptotic scaling as $\textsc{jz-tree}$, but with much larger asymptotic constants. The CPU based $\textsc{scipy-ckdtree}$ turns out more than two orders of magnitude slower than $\textsc{jz-tree}$ and the GPU based kd-tree libraries are more than an order of magnitude slower at $N \gtrsim 10^6$.

This improvement may be largely attributed to several key differences in the tree implementation: (1) The much reduced cost of building a tree in a bottom up approach. (2) The reduced algorithmic cost of a dual (versus one-sided) tree walk and (3) the reduced memory access through warp collaborative evaluation and (4) the improved memory coalescence.

\subsection{Performance across domains} \label{sec:perf_knn_domains}

\begin{figure*}[!t]
    \centering
    \subfloat[]{\includegraphics[width=0.5\textwidth]{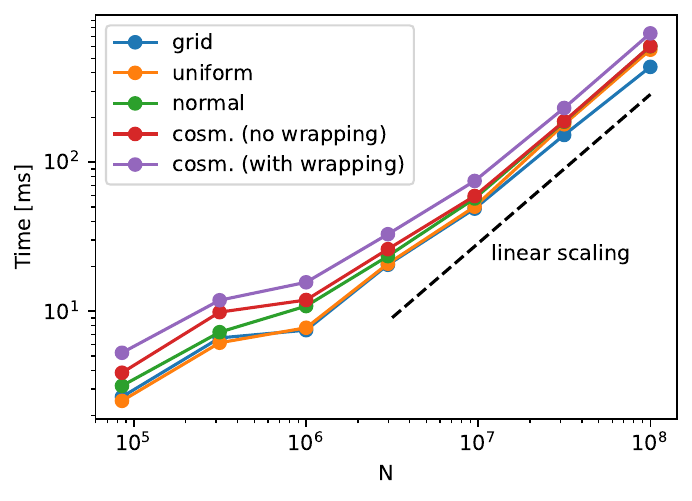}%
    \label{fig:knndistr}}
    \hfil
    \subfloat[]{\includegraphics[width=0.5\textwidth]{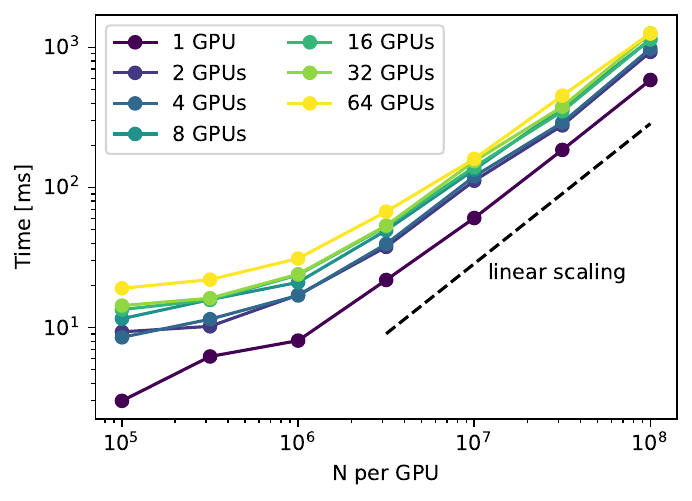}%
    \label{fig:knndevices}}
    \caption{(a) Performance of $\textsc{jz-tree}$ for finding $k=16$ neighbours of different distributions in $d=3$ dimensions. (b) Efficiency scaling for distributed computing for different numbers of GPUs as a function of the number of points \emph{per} GPU. The method scales well across different problem setups and to large problem sizes on multi-GPU.}
    \label{fig:knntests}
\end{figure*}

To demonstrate that the performance benefits are relatively independent of the problem domain, we show in Figure \ref{fig:knntests}(a) performance benchmarks of $\textsc{jz-tree}$ for a variety of different setups. In every case we use query points equal to the source points and look for $k=16$ neighbours in $d=3$ dimensions. The considered scenarios include (1) a uniform grid, (2) the uniform random distribution, (3) a multivariate normal distribution and (4) the final particle distribution from realistic cosmological simulations. The cosmological simulations were run with \textsc{DISCO-DJ} \cite{discodj} in a Planck (2018) cosmology \cite{planck2018} 
with a number of particle-mesh cells and the volume of the box chosen proportionally to the particle count. Specifically we choose the box size as $\sqrt[3]{N} h^{-1} \text{Mpc}$ so that the mass-resolution stays fixed with increasing problem size. For the cosmological simulation we consider two separate scenarios -- one where we appropriately include periodic wrapping in the distance calculation of the kNN -- and one where we don't. For all scenarios we have verified the correctness of the returned neighbour lists against \textsc{scipy-ckdtree}.

It is evident that \textsc{jz-tree} generalizes well over different problem setups with problem-specific performance differences staying well below a factor of two. We note that the most expensive setup -- the cosmological simulation with periodic wrapping -- owes its $20-30\%$ reduction in efficiency primarily to the extra-cost in the wrapping calculation (and not so much to the clustering). If we evaluate the same setup without periodic wrapping, the performance is virtually identical to the uniform random distribution at $N \gtrsim 10^7$.

In Figure \ref{fig:knntests}(b) we evaluate the scaling of the multi-GPU implementation of \textsc{jz-tree} for a self-query of $k=16$ for a uniform random distribution in $d=3$ dimensions. In this test we output the $16$ output indices and radii per point in z-order. 
Importantly, the horizontal axis of the plot shows the number of points per GPU so that e.g. the 64 GPU case with $N \text{ per GPU} = 10^8$ evaluates $1.0 \cdot 10^{11}$ neighbours (16 neighbours for each of $64 \cdot 10^{8}$ points) in about 1.3 seconds.

The method scales well to a large number of GPUs. The biggest drop in the number of evaluations per GPU per second is seen when going from one to two GPUs leading to an increase in evaluation time at $10^8$ from $585 \text{ms}$ up to $928 \text{ms}$ -- close to a factor of two. This increase comes from the additional algorithmic steps that need to be taken for the distributed computing (like rearranging points, sample sort, adjusting domain boundaries and  communication). However, scaling from $2$ to $64$ GPUs exhibits only an additional decrease in efficiency of $30\%$ (928ms for two GPUs versus 1256ms for 64 at $10^8$).

For completeness, we provide additional scaling tests with dimension number, neighbour count and query versus source counts in Appendix \ref{app:knn_details}.
\section{Friends-of-friends clustering} \label{sec:fof}
As a second example algorithm we describe an efficient implementation of FoF clustering here. The implementation follows very closely the previously outlined dual-tree-traversal pattern plus a well known approach for handling linking relations. We have tested it well in $d=2$ and $d=3$ dimensions and for periodic and non-periodic boundary conditions, but the implementation should cleanly generalize to higher dimensional setups as well.

The goal of a FoF algorithm is to find the connected components of a graph where each point is a node and edges exist between every pair of nodes that is closer than the linking length $R_{\mathrm{link}}$ \cite{efstathiou_1985,huchra_1982}. The linking length in cosmological simulations is often chosen relative to the mean separation between points:
\begin{align}
    R_{\mathrm{link}} = \alpha \left(\frac{V}{N}\right)^{1/3}
\end{align}
where $V$ is the volume of the simulation box and $\alpha$ is a parameter that is typically chosen to be $\sim 0.2$, e.g. \cite{lacey_1994}.

\subsection{Implementation} 

The connected components of the FoF graph can be conveniently represented through a pointer $\arr{igroup}$ that is defined per point. If the pointer points to a point itself $\mathrm{igroup}_i = i$, we call $i$ 'a root'. Otherwise, it must point to a point that is of the same group and has a lower index. The root of a point's group can be found by dereferencing the pointer multiple times until it points to itself. All points that have the same root belong to the same group (and vice versa).

The FoF implementation follows the same dual tree walk pattern that is outlined in Algorithm~\ref{alg:knn_dual_walk}. However, in addition to the interaction list, the group pointer $\arr{igroup}^{(p)}$ is carried through the tree-walk and advected from parent to child nodes on every level. It is initialized on the super-node level as a self-pointer. Before the \textsc{NodeToNode} pass, we perform a \textsc{ParentToNode} pass that evaluates
\begin{align}
    \textrm{igroup}_i^{(p)} &= \begin{cases}
        \textrm{spl}^{(p+1)}[\textrm{igroup}_i^{(p+1)}[\text{parent}(i)]] & \text{if linked} \\
        i & \text{else}
    \end{cases}
\end{align}
so that for linked nodes it will point to the first child of the root of its parent node. For unlinked nodes it will simply point to the point itself. A root is considered self-linked if it was linked with any other node or if its diagonal extent is smaller than the linking length.

The node-to-node interaction distinguishes three cases: (1) If both nodes already point to the same root or if $d_{\mathrm{low}} > R_{\mathrm{link}}$, the interaction is discarded. (2) If $d_{\mathrm{up}} \leq R_{\mathrm{link}}$, the other node falls fully inside of the linking length and the nodes are linked together. (3) Otherwise the interaction needs to be evaluated at the child level and is inserted into the interaction list.

When two nodes are linked together, we first find their roots and then update the higher index root to point towards the lower index root -- thereby linking all points in the two groups together. On GPU it is important to protect against data races in this update (between finding the roots and the update, one of the roots may have changed) with atomic compare and swap operations and a repeat on failure.

We first launch a kernel to update $\arr{igroup}$ in this way and afterwards contract the $\arr{igroup}$ relation in a separate kernel. This is simply done by setting every pointer in $\arr{igroup}$ to its root. Finally, we count and insert the interactions that need to be checked on the next level.

The point-point interactions in the leaf-to-leaf kernel only need to distinguish between two scenarios $d > R_{\mathrm{link}}$ -- where the interaction is discarded -- and $d \leq R_{\mathrm{link}}$ -- where the points' groups are linked together. After evaluating these interactions the graph is contracted one final time to obtain a unique label for each group.

The multi-GPU implementation of the FoF requires some additional effort to distinguish between links that can be resolved locally immediately and those that need to be saved to be resolved globally at a later point (involving communication). However, these details are not very relevant with respect to the focus of this paper, so they will be described in Appendix \ref{app:multifof}.

\subsection{Catalogue reduction}

After the group identification, we bring points into group order. That means we perform a stable sort based on the group index, so that the roots of groups remain in z-order with respect to other roots and points in each group form a contiguous block that is internally in z-order. The last group on each device may continue on subsequent devices. Bringing points into group order is useful to make subsequent reduction steps simpler and to make it simple to read the points in different FoF groups separately if the particle data is dumped. 

Finally, we calculate summary statistics like the total mass, the inertia radius, the center of mass position and the the center of mass velocity (if particle velocities were provided as input). This step can be done almost entirely locally, except for a small communication step related to the last/first group on each task. We provide the option to filter the resulting catalogues by a minimum particle count and choose $20$ for this -- as is common in the computational cosmology literature -- in the following performance tests.

\subsection{Performance} \label{sec:fof_performance}

We evaluate the time that is required to obtain the FoF catalogue for the particle distribution from a cosmological simulation (as described in Section~\ref{sec:perf_knn_domains}). This includes all the necessary steps, i.e. sorting, tree building, the tree walk, the reordering into group order and the final reduction steps. However, we don't include disk write time in this benchmark.

For comparison we test against the single CPU FoF implementation of \textsc{hfof}~\cite{Creasey_2018}, the \textsc{MPI} implementation in \textsc{Gadget4}~\cite{springel_2021} and the single GPU implementation of \textsc{jfof}\cite{horowitz_2025}. For \textsc{hfof} we only benchmark the labelling step, since no catalogue reduction is provided -- so results are slightly skewed in its favour. For Gadget4, we read in an hdf5 snapshot that we created with \textsc{DISCODJ} and run only the FoF algorithm. Here, we use the timings that are written into stdout, excluding the initial reading of the input, the initial domain decomposition\footnote{We exclude this, since the input is read initially from a single snapshot onto a single task and is very imbalanced through this until after the first domain decomposition.} and the final writing of the output. We run Gadget once with 1 MPI task and once with 32 MPI tasks on a 32 core node. \textsc{jfof} is the only other pure GPU FoF code that we are aware of and it is a research-level implementation to enable differentible halo finding\cite{horowitz_2025}. It uses $\textsc{jax-kd[CUDA]}$ to iteratively link points together by traversing their neighbour graph. The benchmarks required padding with an additional particle to avoid CUDA memory access errors -- as suggested by the authors.

\begin{figure*}[!t]
    \centering
    \subfloat[]{\includegraphics[width=0.5\textwidth]{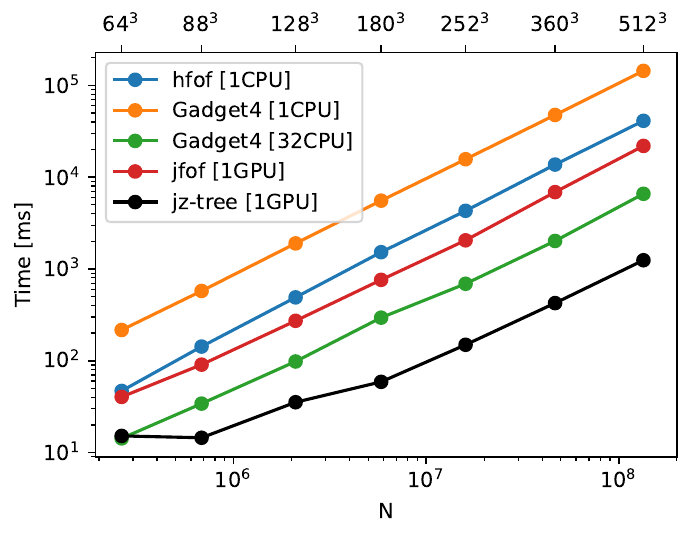}%
    \label{fig:foflibraries}}
    \hfil
    \subfloat[]{\includegraphics[width=0.5\textwidth]{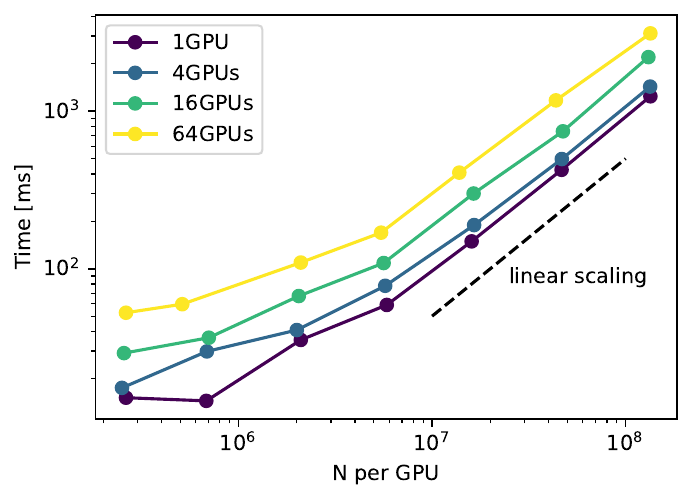}%
    \label{fig:fofdevices}}
    \caption{Performance of the full FoF algorithm of $\textsc{jz-tree-fof}$, including catalogue reduction (a) against other libraries and (b) as a function of the number of devices.}
    \label{fig:foftests}
\end{figure*}

The resulting measurements are found in Figure~\ref{fig:foftests}(a). Similar to the nearest neighbour search, \textsc{jz-tree} scales linearly with the problem size once the GPU is fully saturated $N \gtrsim 10^7$. The performance of $\textsc{jz-tree}$ compares favourably with respect to the alternatives. For $512^3$ the evaluation takes $1.2s$ which is about 5 times faster than Gadget4 with 32 cores (5.3s), 18 times faster than \textsc{jfof} (22s),  66 times faster than hfof (82s) and 116 times faster than Gadget 4 with one core (144s). 

Finally, we show in Figure~\ref{fig:foftests}(b) benchmarks for different GPU counts. The efficiency takes the biggest reduction when jumping from 1 node ($\leq 4$ GPUs) to multiple nodes ($> 4$ GPUs) where the communication becomes less efficient. The most relevant factor here is probably the increased communication latency in the distributed link insertion and contraction steps. However, the efficiency only decreases in total by a factor $2-3$ when scaling from 1 to 64 GPUs, allowing us to calculate FoF group catalouges on $2048^3$ points on 64 GPUs in about $3s$.

\section{Conclusions} \label{sec:conclusions}

Here we have presented a novel approach to construct a plane-based tree hierarchy to enable GPU friendly dual tree walks. Unlike more conventional kd-trees or oct-trees, this tree structure does not partition all of space, has the same depth everywhere and may exhibit a varying number of unequal sized children. It can be constructed in a bottom-up approach with very little additional performance cost after sorted along a Morton z-order curve.

The plane hierarchy allows to implement dual tree walks with good thread collaboration and coalescing memory access patterns. We have demonstrated this on two example applications, nearest neighbour search and FoF clustering -- yielding order of magnitude performance improvements over existing GPU codes with great scaling to distributed computation with large numbers of GPUs.

The presented algorithms are implemented in the \textsc{jz-tree} library, publicly available on GitHub (reference) and PyPI (reference). They can readily be used in HPC simulation schemes that rely on these components like smoothed particle hydrodynamics, self-interacting dark matter simulations and halo finding in cosmological simulations.

Finally, we emphasize that $\textsc{jz-tree}$ forms a suitable framework for developing efficient GPU implementations of other algorithms that rely on tree representations, such as the fast multipole method which we will discuss in an upcoming publication.

\section*{Acknowledgments}
This research was funded in whole or in part by the Austrian Science Fund (FWF) [10.55776/ESP705]. We acknowledge access to the EuroHPC supercomputer LEONARDO, hosted by CINECA (Italy) through the AURELIO call. The authors thank Benjamin Horowitz for help with setting up benchmarks for \textsc{jfof}.

{\appendices
\section{Detailed profiling of kNN} \label{app:knn_details}
In this appendix we evaluate the performance of the nearest neighbour search in $\textsc{jz-tree}$ in dependence on the problem dimension $d$, the neighbour count $k$ and the number of query points. We show the corresponding tests in Figure \ref{fig:knn_dimensions}. Panels (a) and (b) use identical source and query points. Panel (c) varies the number of query points at a fixed number of source points.

The scaling with dimension seems to be close to exponential up to $d=6$ after which it seems to start saturating. At $d=8$ it is still by a factor 10 faster than an evaluation of the same problem with \textsc{FAISS} -- which is quite independent of the dimension number. However, that the performance gap to this brute-force approach is only a factor $10$ makes it seem likely that per query point about $10\%$ of all source points need to be checked. It is quite possible that the scaling with dimension is notably better for more clustered distributions. However, we note that evaluating high dimensional queries requires a very large allocation for the interaction list, limiting the usefulness of our implementation for $d \gg 3$.

\begin{figure*}[!t]
    \centering
    \subfloat[]{\includegraphics[width=0.33\textwidth]{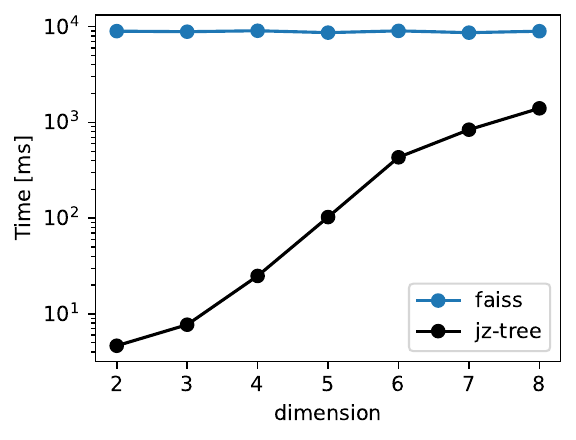}%
    \label{fig:knndim}}
    \hfil
    \subfloat[]{\includegraphics[width=0.33\textwidth]{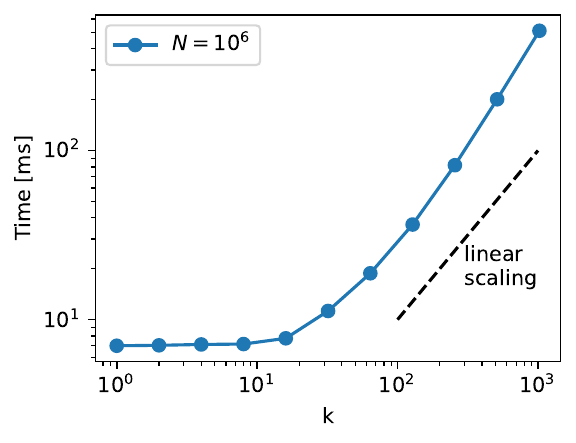}%
    \label{fig:knnk}}
    \hfil
    \subfloat[]{\includegraphics[width=0.33\textwidth]{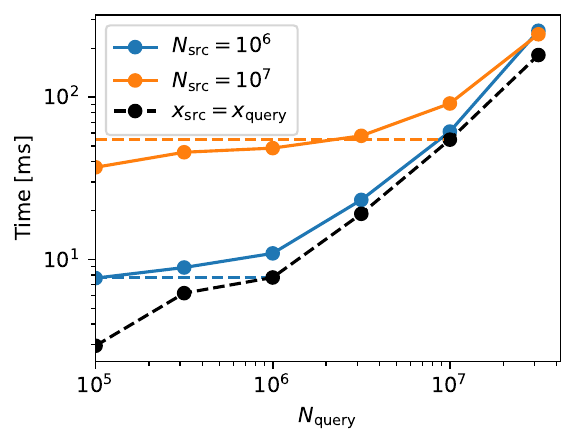}%
    \label{fig:knnquery}}
    \caption{Scaling of the kNN search for $10^6$ points for (a) $k=16$ neighbours with dimension (b) $d=3$ with neighbour number and (c) $k=16$ and $d=3$ and given source count with varying query size.}
    \label{fig:knn_dimensions}
\end{figure*}

In panel (b) we show the scaling with neighbour count. At $k \leq 16$ the evaluation time is almost independent of the neighbour count. This is likely due to our choice of the leaf-size $N_{\mathrm{max}}^{(0)} = 48$, allowing to typically find $O(20)$ neighbours with the same number of leaf-leaf interactions as lower numbers. However, at $k \geq 32$ the evaluation cost scales slightly above linear. The asymptotic super-linearity is likely due to the super-imposed effects of the increased number of traversal kernel launches (requiring $\lceil k/32 \rceil$ kernel launches) plus the increasing size of the volume that needs to be checked.

Finally, in panel (c) we show how the evaluation time depends on the query size for $N_{\mathrm{src}} = 10^6$ and $10^7$ points. Additionally, we show the performance of a self-query as a black line for reference. For large query sizes $N_{\mathrm{query}} \gg N_{\mathrm{src}}$ the algorithm takes slightly more time than a self-query with $N_{\mathrm{query}}$ points. For small query sizes $N_{\mathrm{query}} \ll N_{\mathrm{src}}$ the performance plateaus at a level similar to the time required for a self-query with $N_{\mathrm{src}}$ points. So the performance approximately mirrors the self-query behaviour with $\max(N_{\mathrm{query}}, N_{\mathrm{src}})$ points -- a result of our choice of building the tree based on their joint distribution. 
For scenarios where a large source distribution needs to be evaluated a large number of times with small query distributions this is clearly not optimal. We may consider offering a different approach for this scenario in future releases.

\section{Multi-GPU Friends-of-Friends implementation} \label{app:multifof}
The distributed FoF implementation uses all of the same adaptations that were described in Section \ref{sec:distr_knn} to manage cross-task interactions. However, additional complications arise, because the global root of a node may lie on another rank and may never have appeared in the interaction list. To address this, we first build a local FoF graph that treats every remote node (or point) initially as a root. Additionally we keep track of a set of edges between pairs of points $(\arr{rank}_a, \arr{idx}_a, \arr{rank}_b, \arr{idx}_b)$ that represent links that need to be resolved globally later. Whenever the local updates change the label of a node (or point) with remote origin, we store an edge between the rank and index of the first point in the remote node and its new root.

After the leaf-leaf-interactions have been evaluated, we perform an additional step that resolves the saved links globally. In this step we replace the local $\arr{igroup}$ pointer by a label that includes a rank plus an index pointer. Each edge is sent to the larger involved rank. If the pointed location is (still) a root, the edge can be inserted here by updating that label. Under race conditions we simply resolve the lowest proposed update at the same location and consider the other ones unresolved. If the edge could not be inserted here, we update the larger label with the pointed location and we repeat the procedure (sending the edge to the larger involved rank).

After all links have been inserted in this way, we contract the global graph. This proceeds by requesting for each unique label that points towards a remote rank the label on that rank and index. If the label is different, the local label is updated and the procedure is repeated for those points until all labels are converged.
}

\bibliographystyle{IEEEtran}
\bibliography{references}


 




\vfill

\end{document}